# Heteroepitaxial growth of highly anisotropic Sb$_2$Se$_3$ films on GaAs

Kelly Xiao[1], Virat Tara[2], Pooja D. Reddy[1], Jarod E. Meyer[1], Alec M. Skipper[3], Rui Chen[2], Leland J. Nordin[4,5], Arka Majumdar[2,6], and Kunal Mukherjee[1]*

[1] Department of Materials Science and Engineering, Stanford University, Stanford, CA 94305, USA

[2] Department of Electrical and Computer Engineering, University of Washington, Seattle, WA 98195, USA

[3] Institute for Energy Efficiency, University of California Santa Barbara, Santa Barbara, CA 93106, USA

[4] Department of Materials Science and Engineering, University of Centra Florida, Orlando, FL 32816, USA

[5] CREOL, The College of Optics and Photonics, University of Central Florida, Orlando, FL 32816, USA

[6] Department of Physics, University of Washington, Seattle, WA 98195, USA

*Corresponding Author: kunalm@stanford.edu

The epitaxial integration of anisotropic materials with mainstream cubic semiconductors opens new routes to advanced electronic and photonic devices with directional properties. In this work, we synthesize heteroepitaxial thin films of orthorhombic "quasi-1D" Sb$_2$Se$_3$ on cubic GaAs(001) using molecular beam epitaxy. Traditionally, the synthesis of anisotropic films with low symmetry materials is challenging due to multiple grain orientations that form. On a macroscopic scale, such a film tends towards isotropic properties, even if individual grains possess anisotropic responses. We achieve epitaxial Sb$_2$Se$_3$ grains on pristine homoepitaxial GaAs templates at low temperatures of 180–200 °C. With the Sb$_2$Se$_3$ 1D axis aligned in-plane to GaAs [110] and the primary van der Waals direction lying out-of-plane, we find a birefringence of 0.2 between in-plane orthogonal directions and a giant out-of-plane birefringence greater than 1 at telecom wavelengths. Growth at higher temperatures up to 265 °C yields Sb$_2$Se$_3$ of an unusual in-plane rotated texture that further enhances the in-plane optical index anisotropy to 0.3.

# 1. Introduction

$Sb_2Se_3$ is a layered material in the $V_2$-$VI_3$ chalcogenide family (e.g. $Sb_2Te_3$, $Bi_2Se_3$, $Bi_2Te_3$). Rather than crystallizing in the 2D rhombohedral structure as the prototypical members do, $Sb_2Se_3$ adopts the lower symmetry *Pbnm* orthorhombic structure.[1,2] Orthorhombic $Sb_2Se_3$ is isostructural with stibnite, which hosts covalently chained bonding along only a single "needle" or 1D axis to form parallel ribboned units. Along the other two crystallographic directions, the ribbons loosely pack into a van der Waals ensemble owing to steric effects imparted by the Sb $5s^2$ lone pair.[3] These asymmetric interactions along each primary axis in $Sb_2Se_3$ are the origin of its quasi-1D character. Unsurprisingly, a set of unique optoelectronic and structure-related properties converge in this low-dimensional system, among these favorable electronic and thermal transport along the needle axis,[4] as well as high absorption of the visible wavelengths ($E_g$ = 1.1 – 1.2 eV).[5,6] In light of these compelling qualities, $Sb_2Se_3$ has been proposed as a candidate material for thermoelectric alloys[7,8] and photovoltaic absorbers.[9–14] The inherent 1D nanostructure of this material has additionally inspired a technological focus on its "benign grain boundaries", which in theory are intrinsically passivated and potentially self-healing, enabled by large reconstructive bonding displacements within the open structure.[15,16] Beyond optoelectronic applications, recent efforts are also revisiting the crystalline-amorphous transition of $Sb_2Se_3$ for phase change based reconfigurable photonic devices.[17–19] This reversible transition in the orthorhombic antimonides is rare in that it maintains near-zero losses in the near- to mid-infrared spectral ranges across both defective polycrystalline and amorphous states.[19] Collectively, these applications underscore the broad impact of this material system, pointing to exciting avenues for future research and technological development.

From both fundamental characterization and device standpoints, the properties of $Sb_2Se_3$ are made more intriguing by the strong anisotropy of the crystal,[3,20] and as such, merit investigation of synthesis routes that reliably reproduce and realize such anisotropy to harness its full electronic or optical utility. Focusing on applications in photonics, the very low-loss, high-index, and birefringent character of $Sb_2Se_3$ makes it an ideal candidate for integrated photonics. While prior experimental work on the optical anisotropy of single crystal or bulk $Sb_2Se_3$ is hard to find, first-principles modeling suggests biaxial character with giant birefringence as high as 0.64 between the covalently-bonded 1D axis and primary vdW-bonded direction.[3] This is borne out experimentally in the isostructural bulk $Sb_2S_3$ crystals with birefringence of ~0.9 at 800 nm.[20] Films of naturally anisotropic materials[21] offer a complement to current approaches that use form-birefringence achieved by nanostructuring isotropic materials, which is lossy and challenging to fabricate. Single-crystalline or suitably textured

Sb$_2$Se$_3$ films can offer much needed functionality where birefringence or dichroism is harnessed in integrated components for polarization-sensitive light detection, polarization rotating waveguides, and even higher index contrast in amorphous to single crystalline transitions. As a specific use-case, we note integrated photonics are inherently polarization sensitive, which poses a serious limitation compared to free-space optics. Current efforts in polarization conversion in integrated photonics primarily employ sub-wavelength gratings that inevitably introduce additional loss.[22,23] Therefore, a low-loss anisotropic material integrated on photonic waveguides may help resolve this polarization discrimination.

While dissimilar growth interfaces in heteroepitaxy are energetically unaccommodating, the van der Waals epitaxy growth mode adopted by low-dimensional materials partially relaxes structural and bonding constraints,[24] opening exciting pathways towards integration with conventional single crystal cubic platforms. In this respect, healthy initial progress in heteroepitaxy of several V-VIs such as Bi$_2$Se$_3$, Bi$_2$Te$_3$, and Sb$_2$Te$_3$ on GaAs(001) and Si(111) substrates has been made,[25–29] yet ultra-high vacuum (UHV) thin film synthesis methods for the orthorhombic V-VIs (Sb$_2$Se$_3$ and Sb$_2$S$_3$) remain underdeveloped. With molecular beam epitaxy (MBE), capabilities to monitor surface preparation and produce controlled growth rates may advance tuning of substrate-film interactions. We note demonstrations of heteroepitaxial growth of Sb$_2$Se$_3$ on muscovite mica substrates using vapor transport deposition[30] and on Bi$_2$Se$_3$ epilayers via MBE.[31,32] Both reports are encouraging examples of oriented Sb$_2$Se$_3$ growth, and they target distinct applications for Sb$_2$Se$_3$ in flexible electronics and topological phenomena, respectively. In this study, we aim to address practical challenges involving Sb$_2$Se$_3$ films on surfaces not of hexagonal or trigonal arrangement as has been demonstrated, but on technologically prevalent (001)-type cubic surfaces utilized in photonics. Here, we focus on the commercially accessible GaAs(001) substrate with the objective of exploring the Sb$_2$Se$_3$ synthesis space through growth temperature, Se and Sb beam fluxes, and substrate symmetry parameters. These constitute an initial investigation of whether Sb$_2$Se$_3$ growth is amenable to atomically clean III-V surfaces, and if the film structure can be systematically controlled despite the unorthodox stibnite structure. Sb$_2$Se$_3$ growth on conventional substrates may facilitate device integration beyond photonics; more recently, MBE growth of Sb$_2$Se$_3$ also on GaAs yielded nanostructures potentially useful for electronic applications.[33]

It is important to state that the Sb$_2$Se$_3$ literature has mainly cited two space group conventions (*Pnma* and an equivalent setting *Pbnm*). As evidenced by numerous works dedicated to characterizing and optimizing the orientation of Sb$_2$Se$_3$,[9,13,34–37] explicit and consistent crystallographic indexing of the low symmetry Sb$_2$Se$_3$ material is critical to accurate cross-interpretation.[38] To avoid ambiguity, we reiterate that the *Pbnm* convention ($a$ = 11.63 Å, $b$ = 11.78 Å, $c$ = 3.98 Å, with [001] covalently-bonded chains) has been adopted for conceptual convenience throughout this work.

## 2. Results and Discussion
### 2.1. Nucleation, growth, and morphology

$Sb_2Se_3$ films were prepared using molecular beam epitaxy (MBE) and separate Sb and Se solid sources. The high vapor pressure of Sb and Se require lower growth temperatures to enable sufficient adsorption and subsequent incorporation into the $Sb_2Se_3$ compound. With Sb and Se shutters open, we did not observe changes in reflection high energy electron diffraction (RHEED) until the substrate temperature approached ~280 °C and below, suggesting desorption of the impinging Sb and Se at temperatures above 280 °C. We find successful growth temperatures ($T_g$) for crystalline films were between 180–265 °C. At 150 °C or below, we observe disordered phases of $Sb_2Se_3$ (**Figure S1** and **Figure S2**). The quality of the crystalline films was found to depend on substrate preparation. The first preparation is termed Se-treated GaAs, an epi-ready wafer which undergoes 600 °C thermal deoxidation under a Se overpressure prior to growth; the second is pristine regrown (homoepitaxial) GaAs preserved under an arsenic cap. This surface cap is also thermally desorbed prior to growth, but at much lower temperatures near 400 °C and without an intentional Se overpressure. We refer the reader to the film synthesis methodology for further details on III-V substrate preparation.

Several film morphologies measured by atomic force microscopy (AFM) and corresponding RHEED patterns are compared in **Figure 1**. At substrate temperatures of 265 °C or below, we observe that growth of crystalline $Sb_2Se_3$ directly on Se-treated GaAs forms a rough microstructure. As shown in Figure 1a, protruding and mixed orientation crystallites are plentiful in the film grown on Se-treated GaAs and compromise the overall film quality. In Figure 1b, the chevron pattern RHEED, typical of a rough 3D growth mode, further supports that a Se-treated surface is suboptimal for nucleation and growth. Over the growth duration, the intensity of RHEED slightly dims, suggesting that film quality struggles to improve even after the interface has been overgrown. Misoriented grains are particularly detrimental for $Sb_2Se_3$ film morphology because the crystal habits in stibnites may embody prismatic or bladed forms. This is patently clear in an additional film grown at 265 °C with an increased Sb BEP of $2 \times 10^{-7}$ Torr. (This Sb BEP is 4× greater than the $5 \times 10^{-8}$ Torr BEP otherwise used throughout this work.) These growth conditions promoted inclined shards with poor coalescence across the substrate (**Figure S3**).

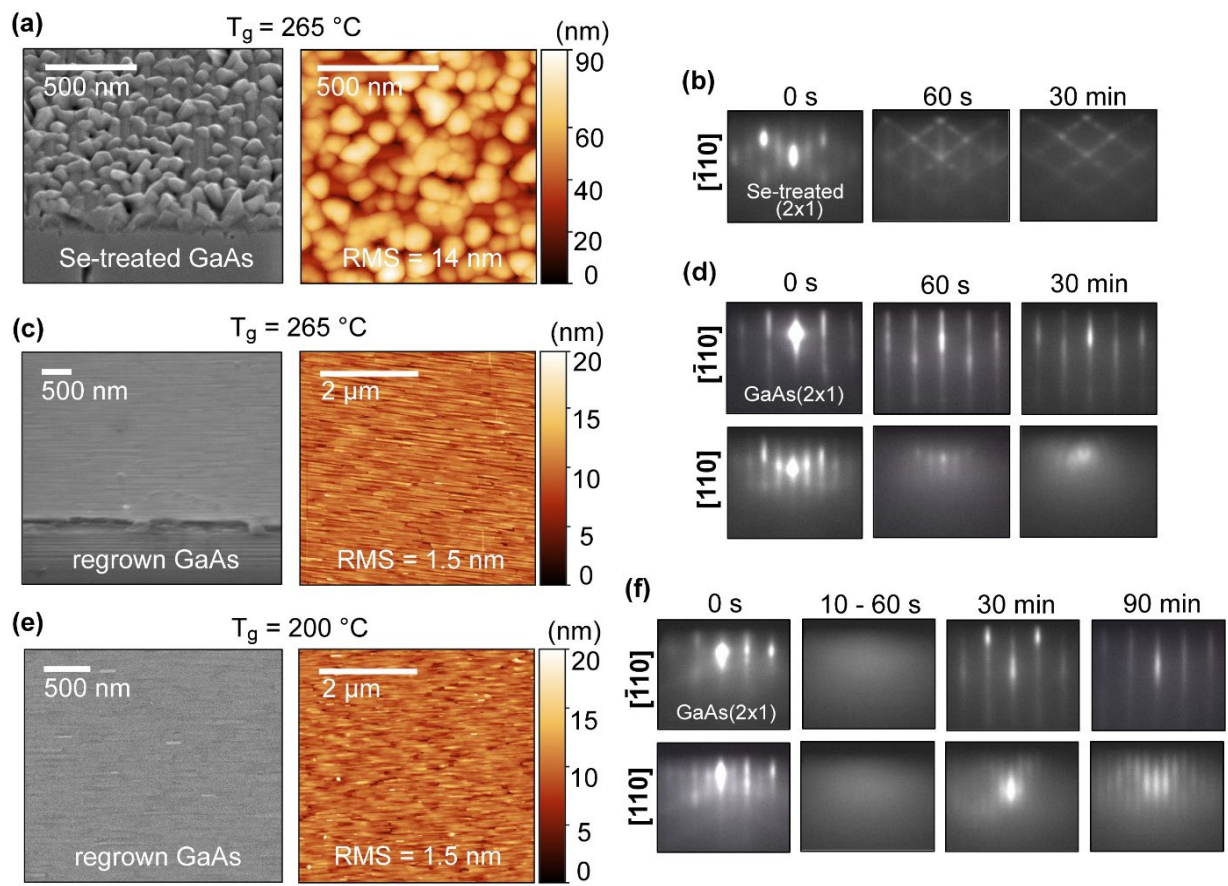

**Figure 1.** (a) 45°-mounted SEM and AFM of Sb$_2$Se$_3$ (265 °C) on Se-treated GaAs and (b) corresponding RHEED pattern. (c) 45°-mounted SEM and AFM of 265 °C Sb$_2$Se$_3$ film on regrown GaAs. A ribbon or needle microstructure due to parallel faceting is visible. (d) Corresponding RHEED pattern for film shown in (c), where streakiness appears along GaAs [$\bar{1}$10] and diffuse inclined streaks gradually develop along [110]. (e) Plan-view SEM and AFM of 200 °C Sb$_2$Se$_3$ on regrown GaAs. A less well-defined ribbon microstructure forms compared to 265 °C. (f) RHEED pattern for 200 °C film. A hazy pattern is observed for initial growth before transitioning to streaks of different periodicities along GaAs [$\bar{1}$10] versus [110].

The structural quality of the films improves significantly on arsenic-capped regrown GaAs templates. We prepared films under the same growth rate of ~0.4 Å/s on regrown GaAs at $T_g$ = 265 °C and 200 °C. Inclined crystallites are significantly reduced, resulting in an overall smoother and continuous film (Figure 1c and 1e). AFM surface topography indicates an order magnitude of reduction in root mean square (RMS) surface roughness from 14 nm to 1.5 nm by switching to regrown GaAs templates. As captured by SEM and AFM, the primary surface feature of the smooth films is parallel faceting, creating a distinct ribbon- or rod-like surface structure. At $T_g$ = 200 °C, these parallel structures become less clearly defined, we suspect due to decreased adatom diffusion at the lower growth temperature. Geometric "rod" surface features were also observed for films grown on mica and Bi$_2$Se$_3$.[30,32] In these reports, domains manifested in parallelogram shapes with 60° or 120° angles, as opposed to the 0°/180° or parallel features we observe.

Improved nucleation can also be immediately observed in the RHEED patterns for films synthesized on regrown GaAs (Figure 1d, 1f). Instead of a chevron pattern previously seen in Figure 1b, $Sb_2Se_3$ RHEED remains streaky throughout the growth period along the GaAs [$\bar{1}$10] direction. Differences emerge in the RHEED pattern following nucleation for these two growth temperatures. At $T_g$ = 265 °C, a faint streaky pattern forms by 60 seconds only along GaAs [110] and further degrades as the film grows thicker. At $T_g$ = 200 °C, upon opening shutters, highly disordered growth first produces a temporary hazy pattern. Following the haze, a two-fold streaky pattern along two primary axes of the orthorhombic cell emerges after approximately 30 seconds to 60 seconds, indicating a transition to longer range order crystalline growth. The [110]-aligned streaks, especially, are initially weak and develop in intensity as the film thickness increases. Their favorable evolution by 90 minutes likely corresponds with gradual improvement in crystallinity along this second axis in $Sb_2Se_3$ as the film increases in thickness. We note that different streak periodicities for the [$\bar{1}$10] and [110] patterns suggest that diffraction arises from real space gratings of shorter and longer periodicity, consistent with the highly dissimilar lattice constants of orthorhombic $Sb_2Se_3$.

The stark contrast in morphology and RHEED between growths on Se-treated and regrown GaAs suggests that a smoother regrown zincblende surface benefits growth quality significantly. Se-treated GaAs may include defects such as Se-terminated dimer rearrangement or Ga-Se byproducts.[39,40] Additionally, Se-treated GaAs develops surface pits due to the high temperatures required for in-situ oxide removal and related outgassing.[41] Pit formation roughening is quite visible in Figure S3 and degrades the seed quality. Lastly, but perhaps most importantly, Se-treated GaAs does not include a so-called "buffer layer", which is used in traditional III-V MBE to smooth out and restore the surface before subsequent epitaxial material is grown. $Sb_2Se_3$ appears to be structurally sensitive to surface defects on GaAs(001) and thus all other films discussed in this study are deposited on pristine regrown GaAs templates to preserve smoothness and crystallinity.

## 2.2 Epitaxial relationship and structural quality

As visually suggested in AFM and SEM, the ribbon surface features of $Sb_2Se_3$ films are oriented in-plane (IP) to the GaAs substrate, prompting us to investigate the structural properties further with X-ray diffraction (XRD). We report two types of crystalline structure in two growth temperature regimes of 230–265 °C (in-plane/IP textured) and 180–200 °C (epitaxial). Symmetric out-of-plane (OP) and phi scans, in addition to several structure schematic diagrams, are shown in **Figure 2** to characterize the grain orientations of $Sb_2Se_3$ on GaAs.

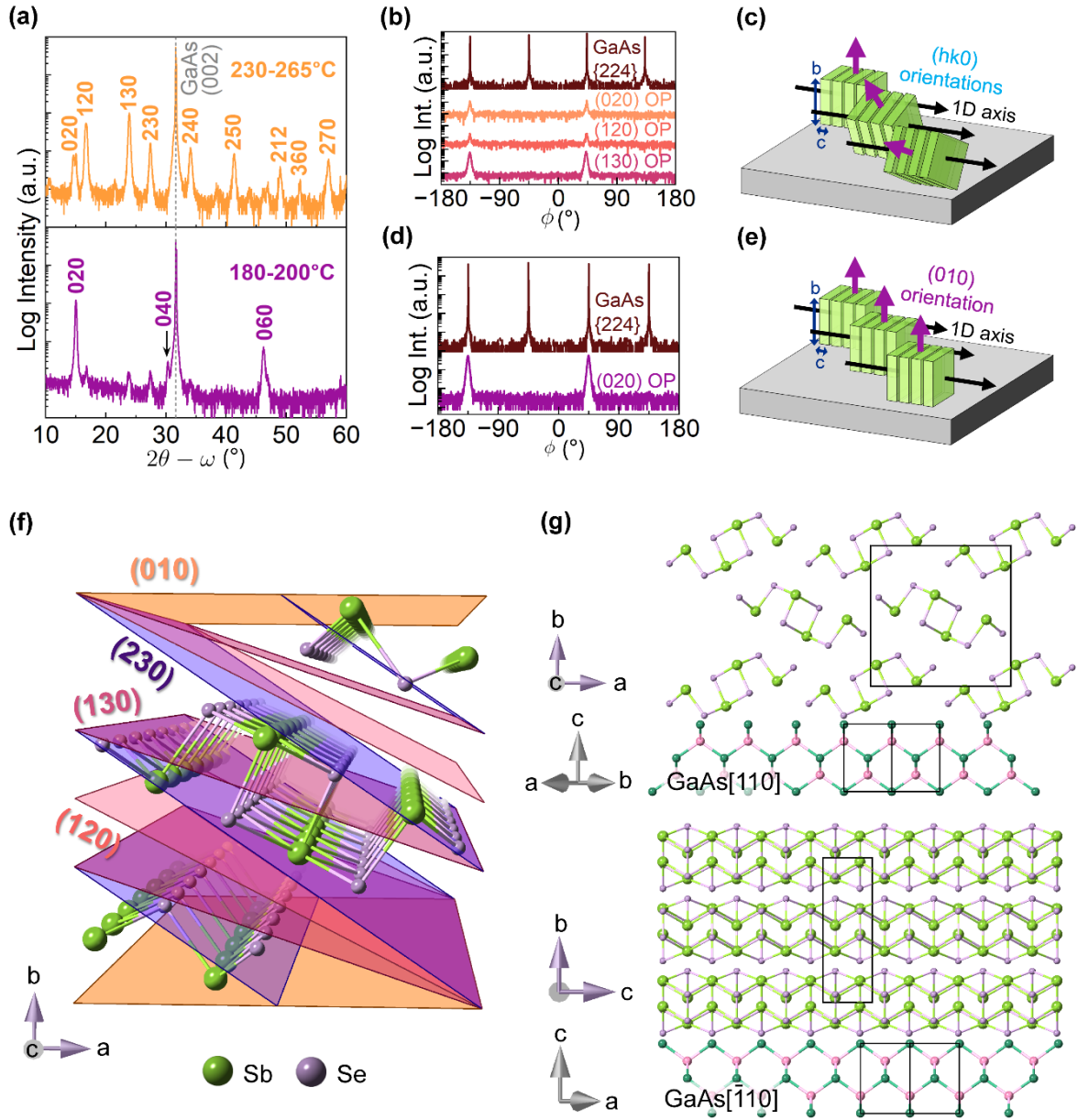

**Figure 2.** (a) Out-of-plane 2θ-ω scans indicate multiple (hk0) orientations are present in $Sb_2Se_3$ synthesized between 230–265 °C and a dominant (010)-orientation is present for 180–200 °C. (b) Phi scans of GaAs{224} and asymmetric $Sb_2Se_3$ reflections for several high volume fraction (hk0)-OP grains in a 230 °C film. The probed $Sb_2Se_3$ asymmetric planes are (120), (020), and (020)-$Sb_2Se_3$, respectively corresponding to the three labeled OP orientations of (020), (120), and (130). The $Sb_2Se_3$ grains show in-plane alignment with GaAs [110]. (c) Schematic diagram of an in-plane textured / (hk0)-oriented film (230–265 °C) showing neighboring grains rotated about their 1D axis. (d) Phi scans of GaAs{224} and the asymmetric $Sb_2Se_3$-(120) reflection in a 200 °C film. $Sb_2Se_3$ still maintains in-plane alignment to GaAs [110]. (e) Schematic diagram of an epitaxial (010)-oriented film (180–200 °C). (f) Perspective view illustration of several (hk0) planes within the $Sb_2Se_3$ unit cell. (hk0) planes are inclined to the horizontal a-c plane and belong to the [001] zone. (g) Epitaxial arrangement of $Sb_2Se_3$ lattice on GaAs (001) viewed along the in-plane GaAs [110] and [1$\bar{1}$0] directions. Unit cells are outlined in black.

First, we characterize the out-of-plane orientations indicated by 2θ-ω scans in Figure 2a. At higher growth temperatures between 230–265 °C, the film contains mostly mixed (hk0)-oriented OP domains, indicated by the presence of both low- and high-index X-ray reflections in the *Pbnm* space group (Figure 2a). Several (hk0) planes are illustrated in Figure 2f. The abundance of both high and low symmetry orientations in Sb$_2$Se$_3$ is an unusual observation; we hypothesize that the predominantly vdW character common to (hk0)-plane terminations is responsible for this phenomenon. In general Sb$_2$Se$_3$ bonding can be decomposed into two components: (i) intrachain interactions (ribbon-forming covalent bonds of length 2.6–2.8 Å) and (ii) interchain interactions (vdW-based with longer interatomic distances).[2] Therefore, the (hk0) planes cut through the ribbons at an oblique angle and primarily terminate the weaker interchain interactions within the Sb$_2$Se$_3$ unit cell.

Intermediate growth temperatures of 180–200 °C produce films favoring a primary (010)-orientation, as indicated by the first allowed (020) and higher order (040) and (060) diffraction peaks in Figure 2a. We believe it is most accurate to categorize this intermediate regime not as entirely independent from the high temperature regime (230–265 °C), but rather as a subtle transition away. Viewed on a logarithmic scale, symmetric 2θ-ω scans reveal that several (hk0) OP orientations present in the $T_g$ = 230 – 265 °C films persist down to 180–200 °C, but their intensities have been dramatically suppressed. This (010)-orientation dominant XRD pattern may point to a more conventional case of vdW materials growth where exposure on a single low energy basal plane is preferred. Several reports have indicated that Sb$_2$Se$_3$ may exhibit such hallmarks of 2D vdW systems despite the quasi-1D nature.[3] The weakest bonding exists along one axis in particular, namely [010] which has the furthest atom-to-atom distance in Sb$_2$Se$_3$.[2] There are several experimental reports on single crystal Sb$_2$Se$_3$ that support this primary (010)-cleavage or (010)-layered behavior.[38,42,43]

As for the in-plane (IP) orientation of Sb$_2$Se$_3$, we find the Sb$_2$Se$_3$ needle axis (*c*-axis) is aligned to the GaAs [110] directions in both films. This crystallographic alignment is consistent with work by Wojnar *et al.*[33] Phi scans of GaAs {224} and two specific asymmetric planes in Sb$_2$Se$_3$ were probed depending on the grain (hk0) OP orientation. For mixed (hk0)-oriented films, we probed the (120)-plane to understand the IP relationship of the (020) OP grains. Likewise, we used the (020)-plane to understand the orientation of (120) and (130) OP grains. These phi scans are shown in relation to the substrate {224} peaks in Figure 2c. We see that two film peaks separated by 180° share the same azimuth as the GaAs [110] IP directions, indicating the IP relationship, Sb$_2$Se$_3$ [001] ∥ GaAs [110], holds for multiple (hk0) OP-oriented grains. An illustrative schematic of these textured films is presented in Figure 2c. In this work, we refer to these films as IP-textured or (hk0)-oriented Sb$_2$Se$_3$. The phi scans also reveal that 180°-rotated twin domains constitute the film. Single crystal orthorhombic Sb$_2$Se$_3$ has a single (020) pole; we have instead observed two 180°-separated peaks arising from the asymmetric (020)-plane in Figure 2b.

The 200 °C epitaxial film, with dominant (010) OP grains, is similarly aligned in-plane to GaAs, as evidenced by a phi scan of the asymmetric (120)-$Sb_2Se_3$ peak (Figure 2d). A schematic of the epitaxial grain structure is depicted in Figure 2e. The epitaxial films are slightly higher crystalline quality than the IP-textured films. Rocking curves (RCs) are large for the IP-textured films, demonstrated by a 1.4° FWHM of the (020) RC and 1.2° FWHM of the (120) RC. The epitaxial films exhibit an out-of-plane (020) RC FWHM = 0.87°, a slight improvement but nonetheless with broadening contributions from low lateral coherence within the film. While these films can be improved structurally, they are distinguished from previous reports of epitaxial $Sb_2Se_3$ because 90°-rotated orthorhombic domains have been largely suppressed on the cubic substrate—resulting in a fully in-plane anisotropic structure not observed for hexagonally-twinned $Sb_2Se_3$ grown on more compatible $Bi_2Se_3$ templates.[32]

Lattice mismatch is a critical parameter that determines the quality of traditional crystalline interfaces and subsequent growth, leading us to consider its potential effects on the orthorhombic-cubic $Sb_2Se_3$/GaAs interface. Epitaxial $Sb_2Se_3$ crystallizes with the *a*- and *c*-axes in-plane, where the highly rectangular *a-c* face is 45°-rotated with respect to the (001) GaAs face. The alignment of orthorhombic $Sb_2Se_3$ on GaAs in parallel and perpendicular view to the covalent chain axis is depicted in Figure 2g. The *a*- and *c*-directions experience different lattice mismatch to the GaAs cubic diagonals. The longer *a*-axis exhibits a 3% mismatch to GaAs [110], derived from a $2d_{SbSe}(001) : 3d_{GaAs}(110)$ ratio. There exists a smaller and thus more favorable 0.5% lattice mismatch between the shorter *c*-axis and GaAs [110] ($2d_{SbSe}(001) : d_{GaAs}(110)$). However, the role of favorable lattice mismatch in mediating the crystallographic relationship between $Sb_2Se_3$ [001] and GaAs [110] is unclear. Instead, we see evidence that substrate symmetry anisotropy plays a large role in influencing the needle direction, as we discuss at the end of this section.

We aim to further understand the transition in growth at the interface for epitaxial films through cross-sectional transmission electron microscopy (XTEM) shown in **Figure 3**. The narrow growth regime in which (010)-oriented epitaxial $Sb_2Se_3$ forms is identifiable by the evolution of the RHEED pattern from hazy to streaky, shown previously in Figure 1f. High-resolution (HR) or phase-contrast XTEM images in Figure 3a show that $Sb_2Se_3$ closer to the interface contains regions of misoriented lattice but is otherwise crystalline. Although hazy RHEED patterns are typically suggestive of amorphous character, in this case, it is possible that the initial growth suffers from such short-range lateral coherency relative to that of the electron beam that coherent diffraction does not occur. RHEED detects coherence on lengths of at least 100 nm.[44,45] We speculate that local clusters which had fortuitously nucleated in the primary vdW (010) orientation were energetically favorable sites for subsequent growth over other grain orientations. An example of such a (010)-oriented region is shown in Figure 3b.

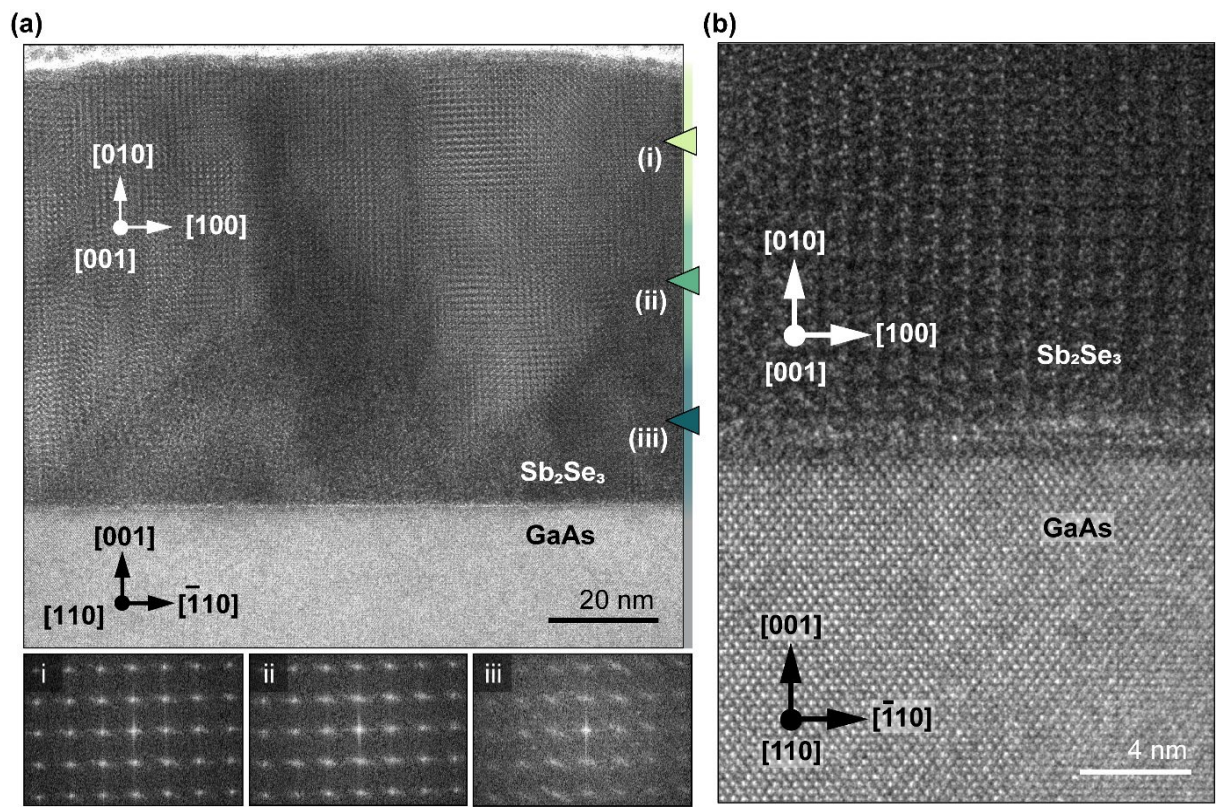

**Figure 3.** (a) HR-TEM of 79 nm 200 °C $Sb_2Se_3$ film imaged along the [110] GaAs zone axis. (i-iii) FFTs at three lateral regions away from the film-substrate interface. The FFT reciprocal lattice exhibits elongated or arced points in region (iii), which is closest to the interface, and collapses into increasingly sharper points by region (ii) and (i). (b) (010)-orientation $Sb_2Se_3$ grain directly nucleated on GaAs.

An alternative explanation for a disorder-to-order RHEED transition was proposed by Shayduk and Braun in MBE-grown cubic $Ge_2Sb_2Te_5$ on GaSb(001).[46] They hypothesized that the initial disordered phase crystallized in epitaxial fashion after sufficient incubation time to seed further epitaxial growth. This growth mechanism is not active in our experiment. Had the epitaxial seed of $Sb_2Se_3$ come from a transformed interface layer, we would expect to resolve lattice fringes signaling improved registry between the interface and the rest of the film, without signatures of overgrowth. In Figure 3a(i)-(iii), we find that fast Fourier transforms (FFTs) taken along three lateral regions gradually further from the interface increasingly sharpen, indicating one contributor to improved structural quality is lower misorientation within the film as the (010)-orientation successfully overgrows. Inspecting the XTEM images further away from the $Sb_2Se_3$/GaAs interface, we find that some regions of the (010)-$Sb_2Se_3$ film produce clearer lattice fringes than others. This additionally points to a wide range of in-plane $Sb_2Se_3$ grain orientations about the GaAs [110] zone axis. The skew-symmetric (120)-$Sb_2Se_3$ RC exhibits a large FWHM of ~1.5°, further supporting the presence of severe twist distortion in the 200 °C film.

The Sb$_2$Se$_3$ films otherwise growing epitaxially or with IP parallel domains requires factors which break the high symmetry of the substrate. We find that the (2 × 1) GaAs surface reconstruction, which likely modulates the interface energy as well as in-plane adatom diffusivity, plays a critical role. We hypothesize anisotropic cation mobility and diffusion across GaAs[47] has promoted the in-plane ordering of Sb$_2$Se$_3$. Directions with longer migration lengths would preferentially promote Sb$_2$Se$_3$ growth along its *c*-axis, as [001] sites are more active and readily available for adatom incorporation. When (2 × 1) GaAs was exchanged with a higher symmetry rocksalt (1 × 1) PbSe(001) template (greater lattice mismatch of 8.0% between Sb$_2$Se$_3$ [001] and PbSe [110]), the film consisted of four 90°-rotated populations instead (**Figure S4**). The relative group VI and V fluxes may also modulate diffusivity anisotropy on GaAs. We have demonstrated that under Sb and Se BEPs of 1 × 10$^{-6}$ Torr and 5 × 10$^{-8}$ Torr, the surface is lightly corrugated from the parallel-faceted grains. However, additional films prepared at 200 °C and lower Se BEPs suggest that the film morphology and in-plane needle alignment suffer (**Figure S5**) without sufficient Se overpressure at 1 × 10$^{-6}$ Torr, highlighting the role of flux tuning to additionally set the epitaxial relationship. This Se BEP is indeed quite large, and we additionally notice that accumulated Se in the chamber negatively impacts the (010) OP orientation upon repeating consecutive 200 °C growths (**Figure S6**). With sufficient time to pump away the volatile Se vapor, we see recovery of epitaxial growth and repeatability of this procedure. Therefore, (i) exploring other low symmetry templates with exaggerated surface anisotropy and (ii) ensuring a low Se environment prior to growth (to preserve a pristine surface) may improve monocrystallinity of Sb$_2$Se$_3$ thin films. This potentially has positive implications for UHV growth on CMOS-compatible cubic substrates that also develop a (2 × 1) reconstruction, such as Si and Ge (001).[48]

## 2.3 Optical anisotropy

The relationship between the structure and optical properties in the Sb$_2$Se$_3$ films are highlighted in this section. We have prepared crystalline-Sb$_2$Se$_3$ films on GaAs that preserve structural anisotropy and therefore are expectedly optically biaxial. The complex refractive index ($n + ik$) across 210–2500 nm obtained from generalized ellipsometry for the IP-textured and epitaxial (010)-Sb$_2$Se$_3$ films are shown in **Figure 4**. We present the in-plane *n* and *k* optical constants of the film approximately along the distinct GaAs [110] and [$\bar{1}$10] directions. Additionally, for epitaxial $Sb_2Se_3$, the out-of-plane index is presented to look for giant birefringence.

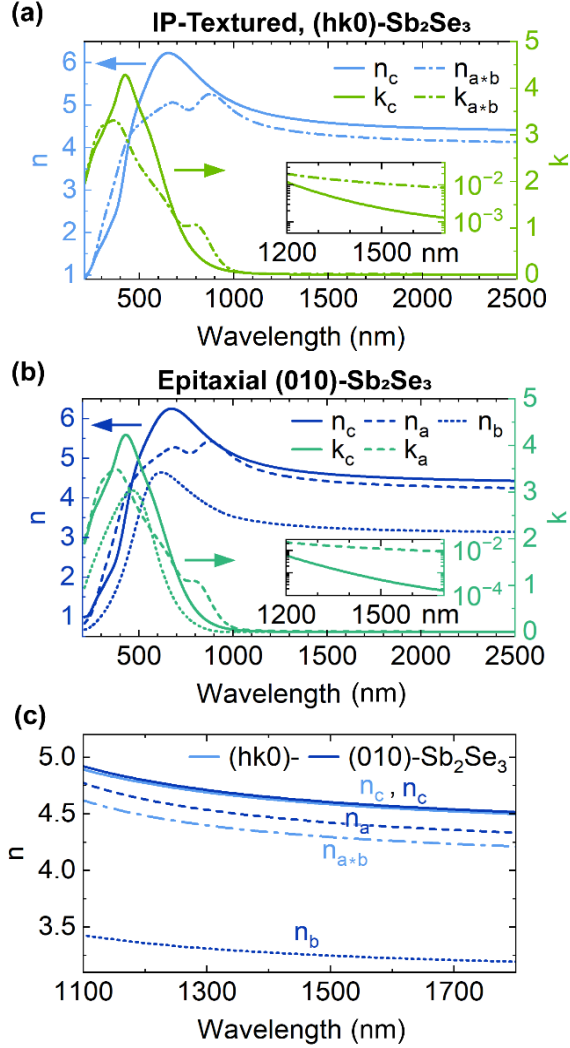

**Figure 4.** Refractive indices and extinction coefficients (n, k) from 210–2500 nm of: (a) in-plane textured / (hk0)-orientation and (b) epitaxial (010)-orientation $Sb_2Se_3$ film. Insets show the extinction coefficients across 1200–1700 nm on a log scale. (c) Comparison of the $n_c$, $n_{a*b}$ refractive indices in (hk0)-oriented $Sb_2Se_3$ versus $n_c$, $n_a$, and $n_b$ in (010)-$Sb_2Se_3$ from 1100 to 1800 nm. The in-plane indices are $n_c$, $n_a$ and the out-of-plane index is $n_b$ for (010)-$Sb_2Se_3$.

The film anisotropy is reflected in the non-zero Mueller matrix off-diagonal terms (**Figure S7** and **Figure S8**). For IP-textured $Sb_2Se_3$, the first set of constants along GaAs [110] ($n_c$, $k_c$) corresponds to polarization along the $Sb_2Se_3$ $c$-axis, and the second set of constants along GaAs [$\bar{1}$10] ($n_{a*b}$, $k_{a*b}$) corresponds to polarization along the $Sb_2Se_3$ $a$- and $b$-directions. (The a∗b subscript denotes weighted average contributions from $a$- and $b$- because of the rotated $Sb_2Se_3$ grains about their $c$-axis.) For epitaxial (010)-$Sb_2Se_3$, the first set of constants are also $n_c$, $k_c$; however, the second set of constants corresponds to polarization primarily along the $a$-axis ($n_a$, $k_a$). The final third set of constants corresponds to polarization in the out-of-plane direction or along the $b$-axis ($n_b$, $k_b$). We caution the reader against ascribing certainty to $n_b$ and $k_b$ in the absorbing region approaching the visible and UV wavelengths. We note that reflection-mode ellipsometry loses sensitivity to out-of-plane orientation when the refracted p-polarization is strongly absorbed in the film, therefore leaving s-polarization, which probes only the in-plane properties.[49] Also, the out-of-plane absorption towards the near-infrared range is assumed to be zero ($k_b = 0$) in our model. We make this assumption based on other experimental reports suggesting Sb$_2$Se$_3$ absorbs minimally below its bandgap.[6,19] In the transparent region, the out-of-plane refractive index fit can be reported with greater confidence.

Looking at the transparent C-band telecom wavelength ($\lambda$ = 1550 nm), IP-textured Sb$_2$Se$_3$ exhibits in-plane indices of $n_c$ = 4.57 and $n_{a*b}$ = 4.28, with a remarkable in-plane birefringence of 0.29. We have not measured the out-of-plane index. Epitaxial (010)-Sb$_2$Se$_3$ exhibits $n_c$ = 4.59, $n_a$ = 4.41, and $n_b$ < 3.5 (Figure 4c)—a giant birefringence exceeding 1 in the *b-c* plane that is among the largest values reported for an epitaxially integrated thin film in the near-infrared. This hierarchy of optical properties arises from its underlying bonding—the needle axis exhibits the highest refractive index (*Pnma*: $n_b$, *Pbnm*: $n_c$), ribbon sheet direction second (*Pnma*: $n_c$, *Pbnm*: $n_a$), and primary vdW direction lowest (*Pnma*: $n_a$, *Pbnm*: $n_b$), in agreement with studies on single crystal Sb$_2$S$_3$ (reported in Schubert *et al.* under the *Pnma* convention[20]). We note that the $n_c$ indices are comparable across these samples as expected, suggesting refracting power is similar in both IP-textured and epitaxial structures for polarization maintained parallel to the needle direction. This considerably expands the temperature growth window for Sb$_2$Se$_3$ on GaAs for photonic applications from epitaxial films to also include the IP-textured films. The $n_{a*b}$ of the IP-textured film is lower than $n_a$ because the index along the primary vdW direction ($n_b$) is lowest in the hierarchy and decreases the weighted index, yielding enhanced IP-anisotropy in these samples compared to the epitaxial film. The birefringence at the O-band ($\lambda$ = 1310 nm) is similar in value to that at 1550 nm.

Near-infrared attenuation is found to be low in these films, as the extinction coefficients $k$ are on the order of $10^{-2}$ and below when the polarization lies within the plane of the film. The confluence of birefringence and high transmission in Sb$_2$Se$_3$ in the near-infrared may be beneficial in devices that require polarization tuning or phase control. Towards the visible range, the in-plane extinction coefficients $k_c$ vs. $k_{a*b}$/$k_a$ are no longer so similar and become of notable magnitude, suggesting dichroism could be an additional functionality for shorter wavelengths, especially around 600–1000 nm.

## 3. Conclusion

We have developed thin films of orthorhombic $Sb_2Se_3$ on cubic GaAs(001) using molecular beam epitaxy. Structurally, these films exhibit preferred alignment of the $Sb_2Se_3$ 1D axis to GaAs [110] at low growth temperatures of 180–265 °C. We demonstrate that these films are low-loss and are optically biaxial and preserve out-of-plane/in-plane optical anisotropy with giant birefringence approaching 1 at telecom bands, and uniquely also an in-plane anisotropy along orthogonal directions approaching 0.3. Our work marks a development towards fully anisotropic $Sb_2Se_3$ films synthesized on technologically relevant cubic templates, a significant departure from previous reports achieving three or six rotational domains that yield in-plane isotropic behavior. We anticipate two important avenues for further study in electronics and photonics. Improvement in the mosaicity and in-plane twist in the epitaxial films will lead to application in a range of minority carrier devices, while transfer of the growth principles derived from our work on GaAs to CMOS-compatible substrates such as Si and Ge would further facilitate integration of $Sb_2Se_3$ and bring exciting orientation-dependent functionalities to existing photonic schemes.

## 4. Experimental Methods

*$Sb_2Se_3$ Thin Film Synthesis*: A Riber Compact 21 (C21) chalcogenide MBE equipped with elemental valved Sb and Se cracker sources was used to grow $Sb_2Se_3$ thin films of thicknesses 72–232 nm on single crystal GaAs(001) substrates. Growth temperatures were read from an optical pyrometer. The beam equivalent pressures (BEPs) of the Sb and Se sources were maintained at $5 \times 10^{-8}$ Torr and $1 \times 10^{-6}$ Torr, respectively, for a growth rate of ~0.4 Å/s.

The majority of the films were grown on MBE-grown homoepitaxial GaAs(001). The regrown GaAs surface was in-situ capped with an amorphous arsenic layer in a separate III-V Veeco Gen III MBE. These substrates were taken out-of-vacuum and indium-bonded to molybdenum platens. The films grown below 150°C were instead prepared on free GaAs substrates, to avoid heating effects that would have otherwise been introduced in the indium-debonding process following growth.)

The substrates were subsequently introduced into the Riber C21 chamber, where the arsenic cap was thermally desorbed and the regrown GaAs substrate was held to stabilize a $(2 \times 4)$ reconstruction. Residual chalcogenide fluxes from beam pressure calibrations or preceding growths within a single day commonly resulted in an immediate conversion to a $(2 \times 1)$ surface. After stabilization of the GaAs surface reconstruction, Sb and Se source shutters were opened to initiate film growth. A Se-treated GaAs surface preparation condition is also presented for comparison, in which the substrate native oxide is thermally desorbed under a Se overpressure of $2 \times 10^{-7}$ Torr prior to initiating growth. The Se flux is supplied due to the absence of an As source in the C21 MBE.

*Structural Characterization*: Film structure was characterized with scanning electron microscopy (SEM), atomic force microscopy (AFM), X-ray diffraction (XRD), and transmission electron microscopy (TEM). Standard plan-view and 45°-mounted scanning electron micrographs of films were acquired using a ThermoFisher Apreo and JEOL JSM-IT500HR SEM. Surface topographical features were analyzed with a Park NX-10 AFM in tapping mode and a NCS15 probe. XRD scans were acquired on a PANanalytical X'Pert Pro MRD instrument with mirror and parallel-plate collimator optics and Ni-filtered Cu-Kα radiation. Cross-sectional lamella of $Sb_2Se_3$ films were milled in a FEI Helios NanoLab 600i DualBeam for high resolution TEM (HR-TEM) 300 keV imaging in a FEI Titan E-TEM.

*Optical Characterization*: Variable angle reflection-mode Mueller matrix ellipsometry measurements between wavelengths of 210–2500 nm were acquired on a J.A. Woollam RC2. To ascertain the anisotropy of the crystalline thin films, three datasets were recorded at multiple azimuthal orientations: the samples were rotated in-plane 0°, 45°, and 90° counterclockwise relative to the $Sb_2Se_3$ 1D axis. The three datasets collected at multiple incidence angles between 55°–75° were used to fit the complex anisotropic refractive index within J.A. Woollam's CompleteEASE analysis software. For crystalline-$Sb_2Se_3$, multiple Gaussian oscillators were employed for fitting of in-plane optical constants. Films of separate thicknesses were used to fit the out-of-plane index under a separate Tauc-Lorentz model.

**Conflicts of Interest**
There are no conflicts to declare.

**Supporting Information**
The data supporting this article have been included as part of the Supplementary Information.

**Acknowledgements**
The authors thank N. Hong from J.A. Woollam for helpful discussions on ellipsometry measurements and fitting. We gratefully acknowledge support for materials synthesis and structural characterization via the NSF CAREER Award under Grant No. DMR-2036520. Part of this work was performed at the Stanford Nanofabrication Facilities (SNF) and Stanford Nano Shared Facilities (SNSF), supported by the National Science Foundation under award ECCS-2026822. A. M., V. T., and R. C. are supported by the DARPA-ATOM program. J. E. M. gratefully acknowledges support from the Tomkat Center for Sustainable Energy's Tomkat Center Graduate Fellow for Translational Research Fellowship.

**References**

1  G. P. Voutsas, A. G. Papazoglou, P. J. Rentzeperis and D. Siapkas, *Z. Kristallogr. - Cryst. Mater.*, 1985, **171**, 261–268.

2  N. W. Tideswell, F. H. Kruse and J. D. McCullough, *Acta Cryst*, 1957, **10**, 99–102.

3  X. Wang, Z. Li, S. R. Kavanagh, A. M. Ganose and A. Walsh, *Phys. Chem. Chem. Phys.*, 2022, **24**, 7195–7202.

4  C. Chen, D. C. Bobela, Y. Yang, S. Lu, K. Zeng, C. Ge, B. Yang, L. Gao, Y. Zhao, M. C. Beard and J. Tang, *Front. Optoelectron.*, 2017, **10**, 18–30.

5  L. R. Gilbert, B. Van Pelt and C. Wood, *J. Phys. Chem. Solids*, 1974, **35**, 1629–1632.

6  C. Wood, Z. Hurych and J. C. Shaffer, *J. Non-Cryst. Solids*, 1972, **8–10**, 209–214.

7  W. M. Yim, E. V. Fitzke and F. D. Rosi, *J Mater Sci*, 1966, **1**, 52–65.

8  H. L. Uphoff and J. H. Healy, *J. Appl. Phys.*, 1963, **34**, 390–395.

9  X. Wen, C. Chen, S. Lu, K. Li, R. Kondrotas, Y. Zhao, W. Chen, L. Gao, C. Wang, J. Zhang, G. Niu and J. Tang, *Nat Commun*, 2018, **9**, 2179.

10 C. Chen, L. Wang, L. Gao, D. Nam, D. Li, K. Li, Y. Zhao, C. Ge, H. Cheong, H. Liu, H. Song and J. Tang, *ACS Energy Lett.*, 2017, **2**, 2125–2132.

11 Y. C. Choi, T. N. Mandal, W. S. Yang, Y. H. Lee, S. H. Im, J. H. Noh and S. I. Seok, *Angew. Chem. Int. Ed.*, 2014, **53**, 1329–1333.

12 M. Leng, M. Luo, C. Chen, S. Qin, J. Chen, J. Zhong and J. Tang, *Appl. Phys. Lett.*, 2014, **105**, 083905.

13 Z. Li, X. Liang, G. Li, H. Liu, H. Zhang, J. Guo, J. Chen, K. Shen, X. San, W. Yu, R. E. I. Schropp and Y. Mai, *Nat. Commun.*, 2019, **10**, 125.

14 G.-X. Liang, Y.-D. Luo, S. Chen, R. Tang, Z.-H. Zheng, X.-J. Li, X.-S. Liu, Y.-K. Liu, Y.-F. Li, X.-Y. Chen, Z.-H. Su, X.-H. Zhang, H.-L. Ma and P. Fan, *Nano Energy*, 2020, **73**, 104806.

15 I. Mosquera-Lois, S. R. Kavanagh, A. Walsh and D. O. Scanlon, *npj Comput. Mater.*, 2023, **9**, 25.

16 K. P. McKenna, *Adv. Electron. Mater.*, 2021, **7**, 2000908.

17 M. Delaney, I. Zeimpekis, D. Lawson, D. W. Hewak and O. L. Muskens, *Adv. Funct. Mater.*, 2020, **30**, 2002447.

18 D. Lawson, D. W. Hewak, O. L. Muskens and I. Zeimpekis, *J. Opt.*, 2022, **24**, 064013.

19 K. Aryana, H. J. Kim, Md. R. Islam, N. Hong, C.-C. Popescu, S. Makarem, T. Gu, J. Hu and P. E. Hopkins, *Opt. Mater. Express*, 2023, **13**, 3277.

20 M. Schubert, T. Hofmann, C. M. Herzinger and W. Dollase, *Thin Solid Films*, 2004, **455–456**, 619–623.

21 S. Niu, G. Joe, H. Zhao, Y. Zhou, T. Orvis, H. Huyan, J. Salman, K. Mahalingam, B. Urwin, J. Wu, Y. Liu, T. E. Tiwald, S. B. Cronin, B. M. Howe, M. Mecklenburg, R. Haiges, D. J. Singh, H. Wang, M. A. Kats and J. Ravichandran, *Nat. Photon.*, 2018, **12**, 392–396.

22 F. Zhang, J. Zheng, Y. Song, W. Liu, P. Xu and A. Majumdar, *OSA Contin.*, 2020, **3**, 560.

23 H. Jin, L. Niu, J. Zheng, P. Xu and A. Majumdar, *Opt. Express*, 2023, **31**, 10684.

24 A. Koma, *Thin Solid Films*, 1992, **216**, 72–76.

25 X. Liu, D. J. Smith, H. Cao, Y. P. Chen, J. Fan, Y.-H. Zhang, R. E. Pimpinella, M. Dobrowolska and J. K. Furdyna, *J. Vac. Sci. Technol. B.*, 2012, **30**, 02B103.

26 Z. Zeng, T. A. Morgan, D. Fan, C. Li, Y. Hirono, X. Hu, Y. Zhao, J. S. Lee, J. Wang, Z. M. Wang, S. Yu, M. E. Hawkridge, M. Benamara and G. J. Salamo, *AIP Adv.*, 2013, **3**, 072112.

27 X. Liu, Y. P. Chen, D. J. Smith, Y.-H. Zhang, C. Liu, M. Z. Hasan, M. Dobrowolska, J. K. Furdyna, J. Fan, H. Cao, T.-L. Wu and R. E. Pimpinella, in *Bismuth-Containing Compounds*, eds. H. Li and Z. M. Wang, Springer New York, New York, NY, 2013, vol. 186, pp. 263–279.

28 Z. Wang and S. Law, *Cryst. Growth Des.*, 2021, **21**, 6752–6765.

29 J. Momand, J. E. Boschker, R. Wang, R. Calarco and B. J. Kooi, *CrystEngComm*, 2018, **20**, 340–347.

30 X. Wen, Z. Lu, L. Valdman, G.-C. Wang, M. Washington and T.-M. Lu, *ACS Appl. Mater. Interfaces*, 2020, **12**, 35222–35231.

31 A. V. Matetskiy, V. V. Mararov, I. A. Kibirev, A. V. Zotov and A. A. Saranin, *J. Condens. Matter Phys.*, 2020, **32**, 165001.

32 Y.-J. Chen, Y.-C. Jhong, P.-Y. Chuang, C.-W. Chong, J.-C.-A. Huang, V. Marchenkov and H.-C. Han, *Chin. J. Phys.*, 2019, **62**, 65–71.

33 P. Wojnar, S. Chusnutdinow, A. Kaleta, M. Aleszkiewicz, S. Kret, J. Z. Domagala, P. Ciepielewski, R. Yatskiv, S. Tiagulskyi, J. Suffczyński, A. Suchocki and T. Wojtowicz, *Nanoscale*, 2024, **16**, 19477–19484.

34 S.-W. Jung, S.-M. Yoon, I.-K. You, B.-G. Yu and K.-W. Koo. *Nanosci. Nanotechnol.*, 2011, **11**, 1569–1572.

35 X. Wen, Z. Lu, B. Li, G.-C. Wang, M. A. Washington, Q. Zhao and T.-M. Lu, *J. Chem. Eng.*, 2023, **462**, 142026.

36 J. Otavio Mendes, A. Merenda, K. Wilson, A. Fraser Lee, E. Della Gaspera and J. Van Embden, *Small*, 2023, 2302721.

37 R. Kondrotas, J. Zhang, C. Wang and J. Tang, *Sol. Energy Mater. Sol. Cells*, 2019, **199**, 16–23.

38 T. D. C. Hobson and K. Durose, *Mater. Sci. Semicond. Process.*, 2021, **127**, 105691.

39 H. Abe, K. Ueno, K. S. Koichiro Saiki and A. K. Atsushi Koma, *Jpn. J. Appl. Phys.*, 1993, **32**, L1444.

40 H. Cheng, J. M. DePuydt, M. A. Haase and J. E. Potts, *J. Vac. Sci. Technol. B.*, 2018, **8**, 181–186.

41 A. Guillén-Cervantes, Z. Rivera-Alvarez, M. López-López, E. López-Luna and I. Hernández-Calderón, *Thin Solid Films*, 2000, **373**, 159–163.

42 N. Fleck, T. D. C. Hobson, C. N. Savory, J. Buckeridge, T. D. Veal, M. R. Correia, D. O. Scanlon, K. Durose and F. Jäckel, *J. Mater. Chem. A*, 2020, **8**, 8337–8344.

43 T. D. C. Hobson, O. S. Hutter, M. Birkett, T. D. Veal and K. Durose, in *2018 IEEE 7th World Conference on Photovoltaic Energy Conversion (WCPEC) (A Joint Conference of 45th IEEE PVSC, 28th PVSEC & 34th EU PVSEC)*, IEEE, Waikoloa Village, HI, 2018, pp. 0818–0822.

44 L. Däweritz, in *Handbook of Surfaces and Interfaces of Materials*, 2001, pp. 351–386.

45 A. Ohtake, *Surf. Sci. Rep.*, 2008, **63**, 295–327.

46 R. Shayduk and W. Braun, *J. Cryst. Growth*, 2009, **311**, 2215–2219.

47 Y. Watanabe, T. Scimeca, F. M. Fumihiko Maeda and M. O. Masaharu Oshima, *Jpn. J. Appl. Phys.*, 1994, **33**, 698.

48 R. J. Hamers, R. M. Tromp and J. E. Demuth, *Phys. Rev. B*, 1986, **34**, 5343–5357.

49 N. Hong, R. A. Synowicki and J. N. Hilfiker, *Appl. Surf. Sci.*, 2017, **421**, 518–528.

# Supporting Information

# **Heteroepitaxial growth of highly anisotropic Sb$_2$Se$_3$ films on GaAs**

Kelly Xiao[1], Virat Tara[2], Pooja D. Reddy[1], Jarod E. Meyer[1], Alec M. Skipper[3], Rui Chen[2], Leland J. Nordin[4,5], Arka Majumdar[2,6], and Kunal Mukherjee[1]*

[1] Department of Materials Science and Engineering, Stanford University, Stanford, CA 94305, USA

[2] Department of Electrical and Computer Engineering, University of Washington, Seattle, WA 98195, USA

[3] Institute for Energy Efficiency, University of California Santa Barbara, Santa Barbara, CA 93106, USA

[4] Department of Materials Science and Engineering, University of Centra Florida, Orlando, FL 32816, USA

[5] CREOL, The College of Optics and Photonics, University of Central Florida, Orlando, FL 32816, USA

[6] Department of Physics, University of Washington, Seattle, WA 98195, USA

*Corresponding Author: kunalm@stanford.edu

## A. Disordered Phase Formation at Growth Temperatures of 150 °C and Below

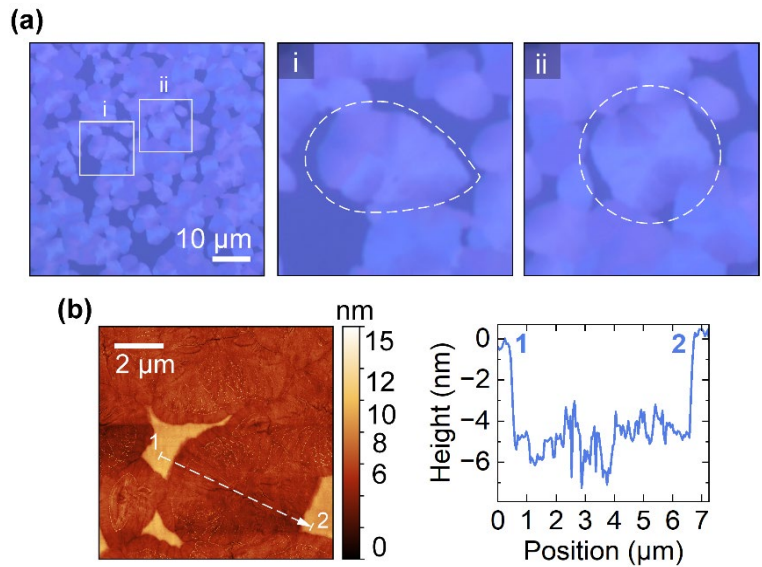

**Figure S1.** $Sb_2Se_3$ spherulites formed during growth at 150 °C on regrown GaAs. (a) Optical microscopy image of the as-grown film, showing spherulites occupy a partial fraction of the amorphous volume. Two variations of spherulite shapes are highlighted: a (i) lenticular and (ii) spherical/radial form. (b) AFM scan of spherulite microstructure in the amorphous matrix. A height profile is shown for a selected line (labeled 1-2) through a spherulite. The elevated sections in the height profile represent the surrounding amorphous matrix.

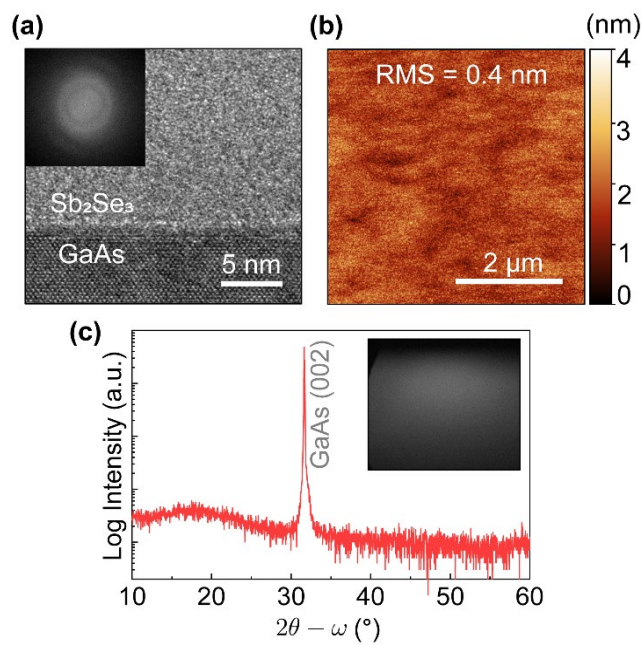

**Figure S2.** Amorphous $Sb_2Se_3$ phase grown below 150 °C on regrown GaAs. (a) (a) HR-TEM image and corresponding ring FFT of amorphous $Sb_2Se_3$ phase. (b) AFM of the amorphous film, showing a smooth surface of RMS roughness = 0.4 nm. (c) XRD out-of-plane $2\theta$-$\omega$ scan of the amorphous film showing no peaks arise from the film, only the substrate. The inset shows the hazy RHEED pattern observed during growth.

Partial spherulite crystallization of an amorphous matrix is observed for growth temperatures of 150 °C. Optical microscopy and AFM images in Figure S1 highlight differences in optical appearance and topographic profiles between the spherulite domains and the amorphous matrix. Below 150 °C, $Sb_2Se_3$ deposits as a majority amorphous phase. Multiple forms of structural characterization shown in Figure S2 (HR-TEM, AFM, XRD, and RHEED) confirm the disordered nature of the film at sub-150 °C growth temperatures.

## B. Effect of Increasing Sb BEP

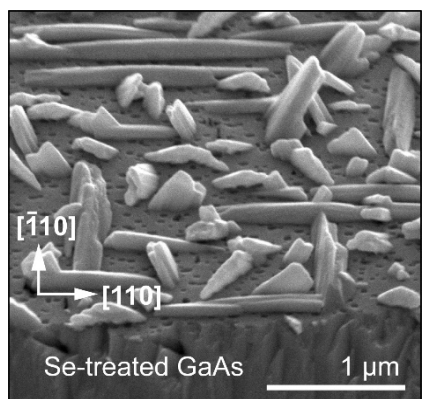

**Figure S3.** 45°-mounted SEM image of Sb$_2$Se$_3$ on Se-treated GaAs, grown with Sb and Se BEP of $2 \times 10^{-7}$ Torr and $1 \times 10^{-6}$ Torr, respectively. There is sparse coverage of Sb$_2$Se$_3$ across the substrate. Pitting of the GaAs substrate is also visible.

We find needle-like growth and very sparse coverage of Sb$_2$Se$_3$ crystals on the GaAs substrate achieved under a Sb and Se BEP of $2 \times 10^{-7}$ Torr and $1 \times 10^{-6}$ Torr, respectively (Figure S3). This film was prepared at 265 °C and a growth duration of 30 minutes, conditions that have produced coalesced films under the same Se BEP and lower Sb BEP of $5 \times 10^{-8}$ Torr. In Figure S3, most crystallites are observed to extend preferentially along the GaAs [110] and [$\bar{1}$10] directions, and a smaller fraction of crystallites grow at an inclined angle to the substrate. Too high a Sb BEP clearly deteriorates the film morphology, as formation of the first and subsequent Sb$_2$Se$_3$ layers have been largely avoided.

## C. Sb$_2$Se$_3$ Films on Rocksalt PbSe Template

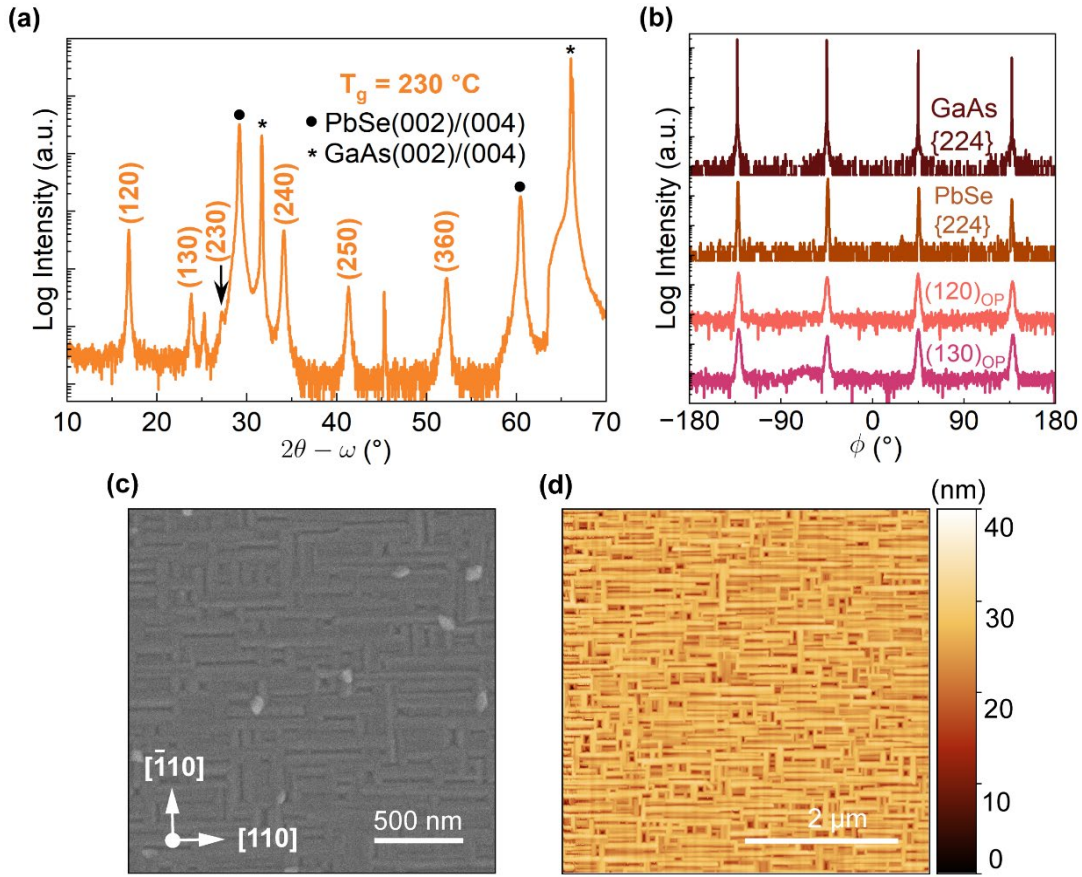

**Figure S4.** (a) XRD 2θ-ω scan showing (hk0)-oriented Sb$_2$Se$_3$ forms on a PbSe(001) template. (120) and (130)-Sb$_2$Se$_3$ are the dominant OP orientations. (b) Phi scan of GaAs{224}, PbSe{224}, and the asymmetric (020)-Sb$_2$Se$_3$ plane for the two dominant (120) and (130)-Sb$_2$Se$_3$ OP orientations. The Sb$_2$Se$_3$ c-axis is aligned in-plane to the four cubic diagonals of PbSe. Film surface microstructure consists of rectangular features observed in (c) SEM and (d) AFM.

A PbSe nucleation and growth sequence adapted from Haidet *et al.*[1] was used to deposit 50 nm of high-quality epitaxial PbSe(001) on GaAs(001), and ~72 nm Sb$_2$Se$_3$ is grown thereafter on PbSe. We hypothesize that a PbSe template may present a more isotropic energy landscape along the [110] and [$\bar{1}$10] directions, creating a weaker bias for diffusion along one direction.

Sb$_2$Se$_3$ growth on PbSe indeed results in 90°-rotated grains aligned to the four cubic diagonals. XRD, SEM, and AFM characterization are shown in Figure S4. In Figure S4a and S4b, out-of-plane (OP) and phi scans indicate that the Sb$_2$Se$_3$ film is primarily (120)- and (130)-oriented OP, and that Sb$_2$Se$_3$ maintains the in-plane relationship Sb$_2$Se$_3$ [001] ∥ PbSe ⟨110⟩. Anti-parallel populations at φ = -135° and 45° constitute the majority fraction in the film, and the other two populations at φ = -45° and 135° show slightly weaker diffraction intensities in the phi scan. With the presence of 90°-rotated domains, neighboring grains impinge on one another at right angles to form the unique rectangular surface features apparent in Figures S4c and S4d. Rectangular features (instead of square) could arise from a deviation from ideal isotropy of the PbSe (001) template.

## D. Effect of Decreasing Se BEP

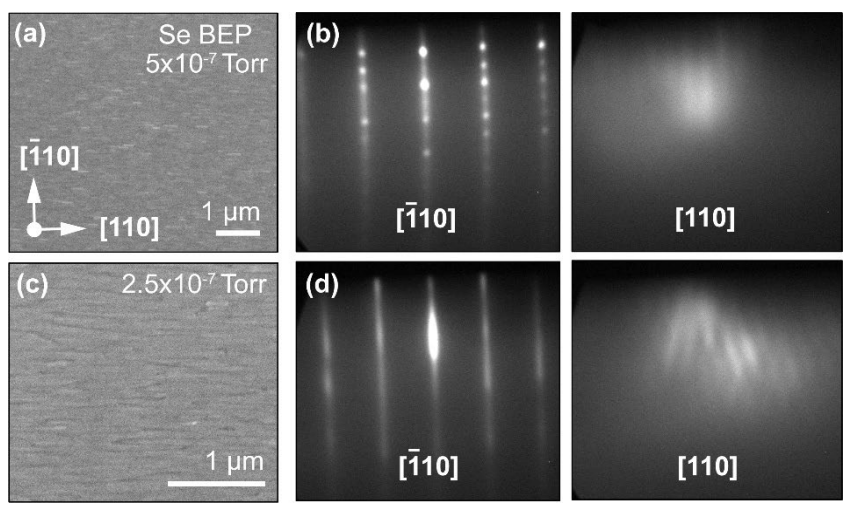

**Figure S5.** Sb$_2$Se$_3$ on regrown GaAs templates at T$_g$ = 200 °C with Sb BEP = 5 × 10$^{-8}$ Torr and variable Se BEP of 5×10$^{-7}$ Torr and 2.5×10$^{-7}$ Torr. (a) A Se/Sb BEP ratio = 10 yields parallel faceting and sub-micron "pill" structures as observed in SEM. (b) Sb$_2$Se$_3$ RHEED patterns along the GaAs [110] and [$\bar{1}$10] directions at 30 mins for the film shown in (a). The [$\bar{1}$10] pattern is spotted along the main streaks and the [110] streaks exhibit a higher periodicity but are noticeably weak. (c) A Se/Sb BEP ratio = 5 results in a slight misangle of the rod-like grains from the horizontal in-plane [110] direction. (d) Sb$_2$Se$_3$ RHEED patterns for the film shown in (c). The [$\bar{1}$10] pattern is streaky, and the [$\bar{1}$10] pattern is complex with inclined intersecting streaks.

Two lower Se flux growth conditions were achieved by lowering the Se BEP (and keeping Sb BEP constant at 5 × 10$^{-8}$ Torr). Surface microstructure and RHEED patterns are shown in Figure S5. The film morphology and crystallographic alignment are found to degrade with lower Se BEP. At Se = 5 × 10$^{-7}$ Torr, grains still appear to primarily align to GaAs [110]; however, the [$\bar{1}$10] RHEED pattern is spotty (Figure S5b). At the lowest explored Se BEP of 2.5 × 10$^{-7}$ Torr, the RHEED pattern along GaAs [110] has degraded from the usual observed vertical streaks (Figure S5d). The streaks are instead doubly inclined and intersecting one another. This film also adopts a "braided" surface appearance, where the ribbon-like grains exhibit a low angle in-plane tilt away from the horizontal GaAs [110] directions (Figure S5c).

# E. Repeatability of Epitaxial Films at $T_g$ = 200 °C

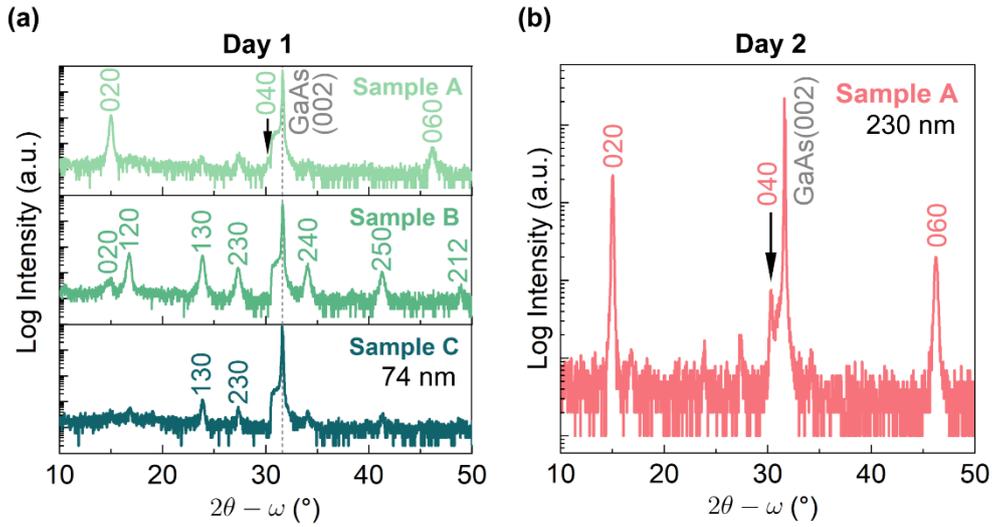

**Figure S6.** (a) Symmetric 2θ-ω scans of 74 nm films grown consecutively on Day 1 (Sample A, B, C) at 200 °C on regrown GaAs, showing the (010) out-of-plane orientation is not consistently achieved after the first growth in the sequence. (b) 2θ-ω scan of first film (~230 nm) grown on Day 2 (Sample A), where the (010) orientation is again observed.

On the same day, thrice repeating the same 200 °C synthesis procedure on regrown GaAs, we find the (010) OP orientation of $Sb_2Se_3$ is no longer maintained after the first growth (Figure S6a). Due to the large Se beam pressure used, we speculate that the residual Se accumulated in the chamber over the duration of consecutive growths alters the substrate surface energy and therefore, the resulting stable crystalline orientations of $Sb_2Se_3$. The desired (010) OP orientation is recovered and obtained on another day after the chamber has returned to baseline conditions (Figure S6b).

## F. Mueller Matrix (MM) for As-Grown Crystalline Films

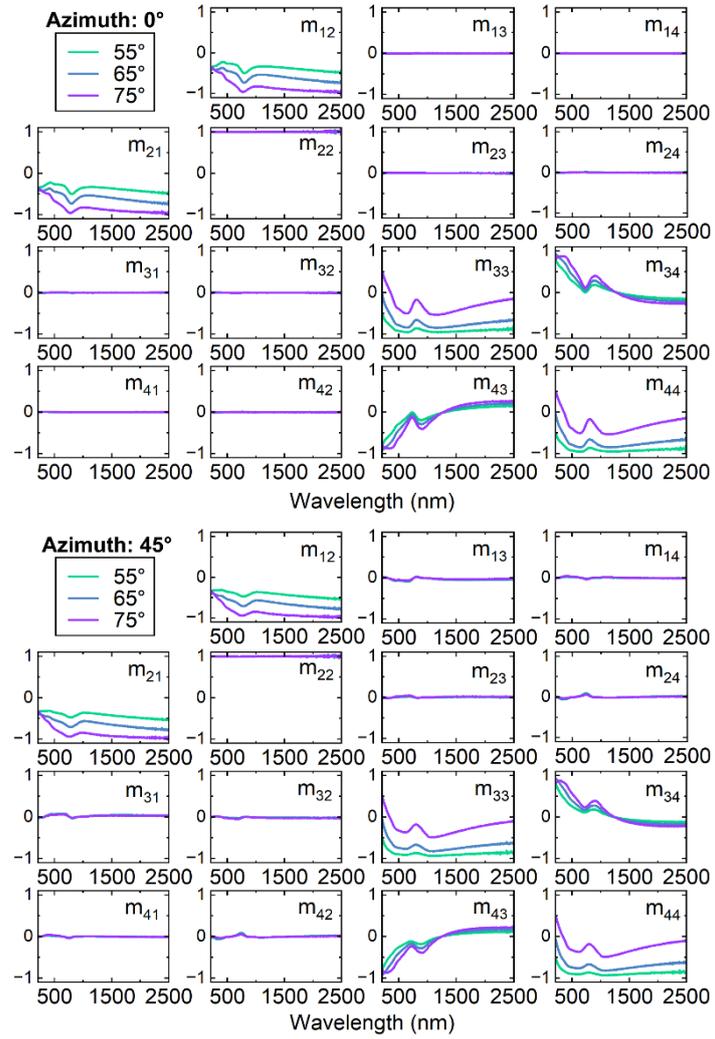

**Figure S7.** Mueller matrix for in-plane textured (hk0)-$Sb_2Se_3$ film across 210–2500 nm for 0° and 45° azimuthal rotation relative to the $Sb_2Se_3$ 1D axis.

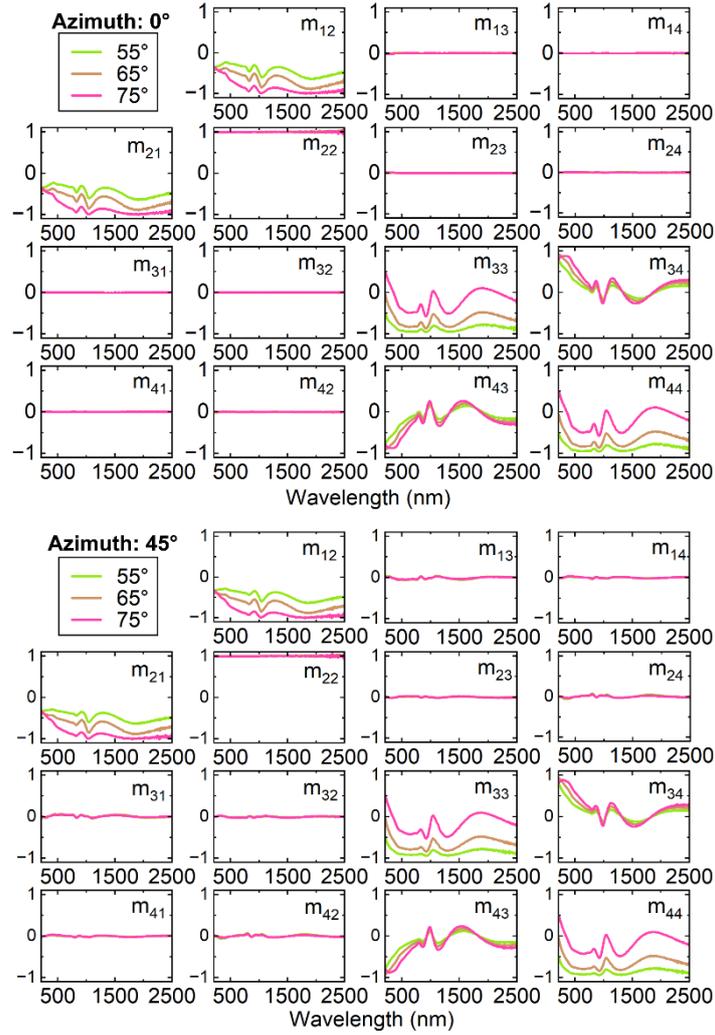

**Figure S8.** Mueller matrix for epitaxial (010)-$Sb_2Se_3$ film across 210–2500 nm for 0° and 45° azimuthal rotation relative to the $Sb_2Se_3$ 1D axis.

Reflection-mode ellipsometry measurements were collected at three variable incidence angles of 55°, 65°, and 75° at approximately 0°, 45°, and 90° azimuthal sample rotations in the counter-clockwise direction relative to the $Sb_2Se_3$ 1D axis direction. Data for 90° azimuthal orientation is not shown as it is represented well by measurements at 0°. At an azimuthal orientation of 45°, non-zero off diagonal blocks in the MM emerge ($m_{13}$, $m_{14}$, $m_{23}$, $m_{24}$ and $m_{31}$, $m_{32}$, $m_{41}$, $m_{42}$), indicating optical anisotropy in the in-plane textured and (010)-epitaxial films (Figure S7 and S8).

## References

1 B. B. Haidet, E. T. Hughes and K. Mukherjee, *Phys. Rev. Materials*, 2020, **4**, 033402.